\documentclass{PoS}
\usepackage{graphicx}
\usepackage{cancel}
\usepackage{amssymb}
\usepackage{textcomp}
\usepackage{amsmath}
\usepackage{bm}
\usepackage{times}
\usepackage{epsfig}
\usepackage{color}
\usepackage{graphics}
\usepackage{setspace}
\usepackage{slashed}

\title{Testing Neutrino Mass Models at the LHC and beyond}

\ShortTitle{Testing Neutrino Mass Models at the LHC and beyond}

\author{P. S. Bhupal Dev$^{1,2}$ \\ 
       $^1$Max-Planck-Institut f\"{u}r Kernphysik, Saupfercheckweg 1, D-69117 Heidelberg, Germany 
$^2$Department of Physics and McDonnell Center for the Space Sciences,
Washington University, St. Louis, MO 63130, USA\\
       E-mail: \email{bdev@wustl.edu}}

\abstract{We briefly review the current status and future prospects of testing the low-scale seesaw models of neutrino mass generation at the energy frontier, with particular emphasis on the ongoing searches at the LHC. }

\FullConference{38th International Conference on High Energy Physics\\
                  3-10 August 2016\\
                 Chicago, USA}

\begin{document}
\section{Introduction}
The observed phenomenon of neutrino oscillations requires at least two of the three active neutrinos to have non-zero masses, which implies physics beyond the Standard Model (SM). A natural way to generate neutrino masses is by breaking the $B-L$ symmetry of the SM, parametrized by the dimension-5 operator $LLHH/\Lambda$~\cite{Weinberg:1979sa}, where $L$ and $H$ are respectively the SM lepton and Higgs doublets, and $\Lambda$ is the new physics scale. There are three tree-level realizations of this operator, commonly known as the type-I~\cite{Minkowski:1977sc}, type-II~\cite{Magg:1980ut} and type-III~\cite{Foot:1988aq} seesaw mechanisms, depending on whether the products of $L$ and $H$ form an $SU(2)_L$ fermion singlet $(L^{\sf T}H)(L^{\sf T}H)/\Lambda$, scalar triplet $(L^{\sf T}\sigma_a L)(H^{\sf T}\sigma_a H)/\Lambda$ or fermion triplet $(L^{\sf T}\sigma_a H)(L^{\sf T}\sigma_a H)/\Lambda$, respectively ($\sigma_a$'s being the usual Pauli matrices).
All of these seesaw mechanisms generically predict lepton number violation (LNV), as well as charged lepton flavor violation (cLFV). These effects might be observable in both energy and intensity frontier experiments, provided the seesaw scale is below a few TeV or so, i.e.~within an experimentally accessible range. There exists a plethora of such low-scale seesaw models; for a review, see e.g. Ref.~\cite{Boucenna:2014zba}. In this  proceedings, we mainly focus on the testable aspects of the simplest seesaw paradigm, viz. the type-I seesaw scenario, though other seesaw scenarios are also briefly mentioned. We summarize the current status and future prospects of the searches for the seesaw messengers at the energy frontier, with particular emphasis on the ongoing Large Hadron Collider (LHC) experiments. For details, see e.g. Ref.~\cite{Deppisch:2015qwa} and references therein.

\section{Type-I seesaw}
This is the simplest extension of the SM for understanding the small neutrino masses. It just requires the addition of SM-singlet Majorana fermions, known as sterile neutrinos $N_\alpha$, to the SM particle content. The relevant piece of the Lagrangian is given by 
\begin{align}
-{\cal L} \ = \ (Y_\nu)_{\ell \alpha} \bar{L}_{\ell}\widetilde{H}N_\alpha+\frac{1}{2}(M_N)_{\alpha\beta}\bar{N}_\alpha^cN_\beta+{\rm H.c.} \, ,
\label{eq:1}
\end{align}  
with $\widetilde H=i\sigma_2 H^*$. The origin of the Majorana mass term $M_N$ can be readily explained in an ultraviolet (UV) completion of seesaw, such as the left-right (LR)  symmetric model~\cite{Pati:1974yy} or $SO(10)$ grand unified theory (GUT)~\cite{Fritzsch:1974nn}. After electroweak symmetry breaking by the Higgs vacuum expectation value (VEV) $v$, Eq.~\eqref{eq:1} leads to a Dirac neutrino mass term $M_D=vY_\nu$ which, together with the Majorana mass $M_N$, induces the tree-level active neutrino masses by the seesaw formula 
\begin{align}
M_\nu \ \simeq \ -M_D M_N^{-1}M_D^{\sf T} \, .
\label{seesaw}
\end{align}
In a bottom-up phenomenological approach, the mass scale of the sterile neutrinos, synonymous with the seesaw scale, is {\em a priori} unknown, and could be anywhere ranging from sub-eV  scale up to the GUT scale $\sim 10^{15}$ GeV. However, theoretical arguments based on the naturalness of the SM Higgs mass of 125 GeV against radiative effects induced by the neutrino loop suggest the seesaw scale to be  below $\sim 10^7$ GeV~\cite{Vissani:1997ys}. Of particular interest to us are TeV-scale seesaw models which are kinematically accessible to the current and foreseeable future collider energies. 

In the minimal type-I seesaw with only the SM gauge group, the active-sterile neutrino mixing parameter $V_{\ell N}\equiv M_DM_N^{-1}$ is also required to be sizable, in addition to a low seesaw-scale,  to get an observable production cross section at colliders~\cite{Datta:1993nm}. 
In the canonical seesaw~\cite{Minkowski:1977sc}, the active-sterile neutrino mixing parameter  is suppressed by the light neutrino mass scale $M_\nu \lesssim 0.1~{\rm eV}$:   
\begin{align}
V_{\ell N} \ \simeq \ \sqrt{\frac{M_\nu}{M_N}} \ \lesssim \ 10^{-6} \sqrt{\frac{100~{\rm GeV}}{M_N}}\; .
\label{canon}
\end{align}
Thus for a TeV-scale seesaw, the experimental effects of the active-sterile neutrino mixing are naively expected to be too small. However, there exist low-scale type-I seesaw models~\cite{Pilaftsis:1991ug, Ibarra:2010xw}, where $V_{\ell N}$ can be sizable due to specific textures of the Dirac and Majorana mass matrices in Eq.~\eqref{seesaw}. These textures can in principle be stabilized by imposing some discrete symmetry in the leptonic sector. But most of these low-scale seesaw scenarios require the sterile neutrinos having large mixing with the active sector to be quasi-Dirac, thus suppressing all LNV effects, except when the mass splitting between the pseudo-Dirac sterile neutrino pair is comparable to their decay width, which could lead to a resonant enhancement of the LNV amplitude~\cite{Bray:2007ru}. On the other hand, if the heavy sterile neutrinos have additional gauge interactions, e.g.~when they are charged under the $SU(2)_R$ gauge group in LR seesaw models, one could get large LNV signals at colliders, irrespective of their mixing with the active neutrino sector~\cite{Keung:1983uu}. Note that the LNV effects involving only the electron flavor are strongly constrained from neutrinoless double beta decay searches~\cite{Ibarra:2010xw, Lopez-Pavon:2015cga}, but those involving the muon and/or tau flavors could still allow for an observable signal at colliders, depending on the model construction.

Another natural realization of low-scale seesaw with potentially large active-sterile neutrino mixing is the inverse seesaw mechanism~\cite{Mohapatra:1986aw}, which introduces two sets of SM-singlet pseudo-Dirac fermions $(N,S)$ with only a small Majorana mass term $\mu_S$ for $S$. In this case, the magnitude of the light neutrino mass can be decoupled from the heavy neutrino mass scale: 
\begin{align}
M_\nu \ \simeq \ (M_DM_N^{-1})\mu_S (M_D M_N^{-1})^{\sf T} \, ,
\end{align} 
thus allowing for large active-sterile mixing even for TeV-scale seesaw without any fine-tuning: 
\begin{align}
\label{eq:thetainvseesaw}
		V_{\ell N} \ \simeq \ \sqrt{\frac{M_\nu}{\mu_S}} \ \approx \ 10^{-2}\sqrt{\frac{1~\text{keV}}{\mu_S}} \; . 
\end{align}
Note that the smallness of $\mu_S$ is technically natural, i.e. in the limit of $\mu_{S}\to 0$, lepton   number  symmetry   is  restored   and  the light  neutrinos
 are exactly massless  to all  orders in  perturbation theory. However, the smallness of $\mu_S$ implies that all LNV effects will be suppressed. This could be partially overcome by introducing a Majorana mass term $\mu_N$ for the $N$ fermions~\cite{Dev:2012sg}. 

The hadron collider experiments can {\em simultaneously} probe the Majorana nature of the neutrinos and the active-sterile neutrino mixing parameters through the ``smoking gun" LNV signature of same-sign dilepton plus two jets: $pp\to N\ell^\pm \to \ell^\pm \ell^\pm jj$~\cite{Datta:1993nm}. This is in contrast with   the complementary low-energy probes at the intensity frontier~\cite{deGouvea:2013zba} which are mostly sensitive to only one aspect, e.g. neutrinoless double beta decay for the Majorana nature and cLFV searches for the active-sterile neutrino mixing. The current direct search limits using the same-sign dilepton channel at $\sqrt s=8$ TeV LHC range from $|V_{\ell N}|^2 \lesssim 10^{-2}-1$ (with $\ell=e,\mu$) for $M_N=100-500$ GeV~\cite{Khachatryan:2015gha}. These limits could be improved by roughly an order of magnitude and extended for heavy neutrino masses up to a TeV or so with the $\sqrt s=14$ TeV LHC and further improvements by another order of magnitude are possible at the proposed 100 TeV $pp$ collider~\cite{Deppisch:2015qwa}. It is worth emphasizing that the $W\gamma$ vector boson fusion processes~\cite{Dev:2013wba} become increasingly important at higher center-of-mass energies and/or higher masses, and must be taken into account, along with the usual Drell-Yan production mechanism via $s$-channel $W$ boson so far considered in the experimental analyses.  

To test seesaw models with suppressed LNV effects, as well as to distinguish between Dirac and Majorana nature of the sterile neutrinos, one should also study the opposite-sign dilepton signal $pp\to N\ell^\pm \to \ell^\pm \ell^\mp jj$ and its ratio with the same-sign signal~\cite{Dev:2015pga}. Although the opposite-sign dilepton signal suffers from a huge SM background, mostly from $Z\to \ell^+ \ell^-$ decays, one can exploit specific kinematic features to get a good signal-to-background ratio~\cite{Khachatryan:2014dka}. Another promising channel is the trilepton mode $pp\to N\ell^\pm \to \ell^\pm \ell^\mp \ell^\pm + \slashed{E}_T$\cite{delAguila:2008cj}, which has a relatively smaller cross section, but a smaller SM background as well. In addition, there exist indirect signals for electroweak-scale sterile neutrinos which can probe large active-sterile neutrino mixing, irrespective of their Majorana nature, such as anomalous Higgs~\cite{Dev:2012zg} and $W/Z$ decays~\cite{Blondel:2014bra}. For sterile neutrinos lighter than the $W$ boson, one could also look for displaced vertex signatures~\cite{Helo:2013esa}. GeV-scale sterile neutrinos are also constrained at LHCb from searches in rare $B$-meson decays~\cite{Aaij:2014aba}. In future, this mass range can be more effectively probed in beam dump experiments, such as SHiP~\cite{Alekhin:2015byh}. 

A future lepton collider, such as ILC or FCC-ee, will provide better sensitivity up to sterile neutrino masses very close to its kinematic threshold~\cite{Antusch:2015mia}, mainly due to its relatively cleaner environment, as compared to hadron colliders. See also Ref.~\cite{Mondal:2015zba} for sterile neutrino studies at a future electron-proton collider, such as LHeC. For a summary of all relevant constraints and future prospects for sterile neutrinos over MeV-TeV mass range, see Refs.~\cite{Deppisch:2015qwa, Atre:2009rg}. 

\section{Type-II seesaw}
Unlike the sterile neutrinos in the minimal type-I seesaw which, being SM gauge singlets, can only communicate with the SM sector through their mixing with active neutrinos, the type-II seesaw messenger, being an $SU(2)_L$-triplet scalar $(\Delta^{++},\Delta^+,\Delta^0)$, can be {\it directly} produced at the LHC via its gauge interactions. The smoking gun collider signal in this case would be the detection of the doubly-charged scalars with LNV interactions. The most promising channels at the LHC are $pp\to Z^*/\gamma^*\to \Delta^{++}\Delta^{--}\to \ell^+\ell^+\ell^-\ell^-$ and $pp\to W^\pm \to \Delta^{\pm}\Delta^{\mp}\to \ell^\pm\ell^\pm \ell^\mp \nu_\ell$ using which lower limits on the triplet mass of up to 600 GeV have been derived at the LHC~\cite{ATLAS:2014kca}. These limits could be marginally improved up to about 800 GeV at the $\sqrt s=14$ TeV LHC, whereas a future 100 TeV $pp$ collider could probe up to 5 TeV or so~\cite{Dev:2016dja}, depending on the model parameters.  

\section{Type-III seesaw} 
Similar to the type-II seesaw case, the type-III seesaw messenger, being an $SU(2)_L$-triplet fermion $(\Sigma^+,\Sigma^0,\Sigma^-)$, can be directly produced at the LHC via its gauge interactions. The most promising production channels at the LHC are $pp\to W^{\pm *}\to \Sigma^\pm \Sigma^0$ and $pp\to Z^*/\gamma^*/h^* \to \Sigma^+\Sigma^-$, which lead to multi-lepton final states. Depending on the theoretical scenario considered, the $\sqrt s=8$ TeV LHC data has excluded fermion triplets of mass up to 540 GeV at 95\% CL~\cite{Aad:2015cxa}. These limits could be improved up to about 1 TeV at the $\sqrt s=14$ TeV LHC. 

\section{Conclusion}
Neutrino oscillations have provided us with the first (and so far only) conclusive experimental evidence of physics beyond the SM. Therefore, it is very important to explore the experimental signatures  of  neutrino  mass  models, which might lead to some crucial insights into the underlying new  physics. A simple theoretical paradigm for neutrino masses is the seesaw mechanism, which also provides potentially attractive solutions to other outstanding puzzles like dark matter and baryon asymmetry of the universe. We have briefly reviewed the current status and future prospects of the direct searches for various seesaw messengers at the energy frontier, and show that colliders offer an ideal testing ground for low-scale seesaw models, which is complementary to the low-energy probes at the intensity frontier.  




\begin{thebibliography}{99}

\bibitem{Weinberg:1979sa} 
  S.~Weinberg,
  Phys.\ Rev.\ Lett.\  {\bf 43}, 1566 (1979).

 
\bibitem{Minkowski:1977sc} 
  P.~Minkowski,
  Phys.\ Lett.\ B {\bf 67}, 421 (1977); 
  R.~N.~Mohapatra and G.~Senjanovi\'c,
  Phys.\ Rev.\ Lett.\  {\bf 44}, 912 (1980); 
%
  T.~Yanagida,
  Conf.\ Proc.\ C {\bf 7902131}, 95 (1979); 
%
  M.~Gell-Mann, P.~Ramond and R.~Slansky,
  Conf.\ Proc.\ C {\bf 790927}, 315 (1979); 
  S.~L.~Glashow,
  NATO Sci.\ Ser.\ B {\bf 61}, 687 (1980).

\bibitem{Magg:1980ut} 
J.~Schechter and J.~W.~F.~Valle,
  Phys.\ Rev.\ D {\bf 22}, 2227 (1980); 
T.~P.~Cheng and L.~F.~Li,
  Phys.\ Rev.\ D {\bf 22}, 2860 (1980); 
   G.~Lazarides, Q.~Shafi and C.~Wetterich,
  Nucl.\ Phys.\ B {\bf 181}, 287 (1981); 
R.~N.~Mohapatra and G.~Senjanovi\'{c},
  Phys.\ Rev.\ D {\bf 23}, 165 (1981).

\bibitem{Foot:1988aq} 
  R.~Foot, H.~Lew, X.~G.~He and G.~C.~Joshi,
  Z.\ Phys.\ C {\bf 44}, 441 (1989).


\bibitem{Boucenna:2014zba} 
  S.~M.~Boucenna, S.~Morisi and J.~W.~F.~Valle,
  Adv.\ High Energy Phys.\  {\bf 2014}, 831598 (2014). 

\bibitem{Deppisch:2015qwa} 
  F.~F.~Deppisch, P.~S.~B.~Dev and A.~Pilaftsis,
  New J.\ Phys.\  {\bf 17}, 075019 (2015). 

\bibitem{Pati:1974yy} 
  J.~C.~Pati and A.~Salam,
  Phys.\ Rev.\ D {\bf 10}, 275 (1974); 
  R.~N.~Mohapatra and J.~C.~Pati,
  Phys.\ Rev.\ D {\bf 11}, 2558 (1975); 
  G.~Senjanovi\'{c} and R.~N.~Mohapatra,
  Phys.\ Rev.\ D {\bf 12}, 1502 (1975).

\bibitem{Fritzsch:1974nn} 
  H.~Fritzsch and P.~Minkowski,
  Annals Phys.\  {\bf 93}, 193 (1975).

\bibitem{Vissani:1997ys} 
  F.~Vissani,
  Phys.\ Rev.\ D {\bf 57}, 7027 (1998); 
  J.~D.~Clarke, R.~Foot and R.~R.~Volkas,
  Phys.\ Rev.\ D {\bf 91}, 073009 (2015); 
G.~Bambhaniya {\it et al.}, 
  arXiv:1611.03827 [hep-ph].

\bibitem{Datta:1993nm} 
  A.~Datta, M.~Guchait and A.~Pilaftsis,
  Phys.\ Rev.\ D {\bf 50}, 3195 (1994); 
  O.~Panella, M.~Cannoni, C.~Carimalo and Y.~N.~Srivastava,
  Phys.\ Rev.\ D {\bf 65}, 035005 (2002); 
  T.~Han and B.~Zhang,
  Phys.\ Rev.\ Lett.\  {\bf 97}, 171804 (2006); 
  F.~del Aguila, J.~A.~Aguilar-Saavedra and R.~Pittau,
  JHEP {\bf 0710}, 047 (2007). 


\bibitem{Pilaftsis:1991ug} 
  A.~Pilaftsis,
  Z.\ Phys.\ C {\bf 55}, 275 (1992); 
%
  J.~Gluza,
  Acta Phys.\ Polon.\ B {\bf 33}, 1735 (2002); 
  J.~Kersten and A.~Y.~Smirnov,
  Phys.\ Rev.\ D {\bf 76}, 073005 (2007);    
  Z.~z.~Xing,
  Prog.\ Theor.\ Phys.\ Suppl.\  {\bf 180}, 112 (2009); 
  X.~G.~He, S.~Oh, J.~Tandean and C.~C.~Wen,
  Phys.\ Rev.\ D {\bf 80}, 073012 (2009); 
  R.~Adhikari and A.~Raychaudhuri,
  Phys.\ Rev.\ D {\bf 84}, 033002 (2011);
  F.~F.~Deppisch and A.~Pilaftsis,
  Phys.\ Rev.\ D {\bf 83}, 076007 (2011);  
  M.~Mitra, G.~Senjanovic and F.~Vissani,
  Nucl.\ Phys.\ B {\bf 856}, 26 (2012);  
  C.~H.~Lee, P.~S.~B.~Dev and R.~N.~Mohapatra,
  Phys.\ Rev.\ D {\bf 88}, 093010 (2013). 

\bibitem{Ibarra:2010xw} 
  A.~Ibarra, E.~Molinaro and S.~T.~Petcov,
  JHEP {\bf 1009}, 108 (2010); 
  Phys.\ Rev.\ D {\bf 84}, 013005 (2011). 

\bibitem{Bray:2007ru} 
  S.~Bray, J.~S.~Lee and A.~Pilaftsis,
  Nucl.\ Phys.\ B {\bf 786}, 95 (2007).  




\bibitem{Keung:1983uu} 
  W.~Y.~Keung and G.~Senjanovi\'c,
  Phys.\ Rev.\ Lett.\  {\bf 50}, 1427 (1983); 
  A.~Ferrari {\it et al}, 
  Phys.\ Rev.\ D {\bf 62}, 013001 (2000); 
  M.~Nemevsek, F.~Nesti, G.~Senjanovic and Y.~Zhang,
  Phys.\ Rev.\ D {\bf 83}, 115014 (2011); 
  S.~P.~Das, F.~F.~Deppisch, O.~Kittel and J.~W.~F.~Valle,
  Phys.\ Rev.\ D {\bf 86}, 055006 (2012); 
  J.~A.~Aguilar-Saavedra and F.~R.~Joaquim,
  Phys.\ Rev.\ D {\bf 86}, 073005 (2012); 
  C.~Y.~Chen, P.~S.~B.~Dev and R.~N.~Mohapatra,
  Phys.\ Rev.\ D {\bf 88}, 033014 (2013); 
  J.~N.~Ng, A.~de la Puente and B.~W.~P.~Pan,
  JHEP {\bf 1512}, 172 (2015); 
  P.~S.~B.~Dev, D.~Kim and R.~N.~Mohapatra,
  JHEP {\bf 1601}, 118 (2016). 

\bibitem{Lopez-Pavon:2015cga} 
  J.~Lopez-Pavon, E.~Molinaro and S.~T.~Petcov,
  JHEP {\bf 1511}, 030 (2015); 
  E.~Fernandez-Martinez, J.~Hernandez-Garcia, J.~Lopez-Pavon and M.~Lucente,
  JHEP {\bf 1510}, 130 (2015). 


\bibitem{Mohapatra:1986aw} 
  R.~N.~Mohapatra and J.~W.~F.~Valle,
  Phys.\ Rev.\ D {\bf 34}, 1642 (1986).



\bibitem{Dev:2012sg} 
  P.~S.~B.~Dev and A.~Pilaftsis,
  Phys.\ Rev.\ D {\bf 86}, 113001 (2012); 
 Phys.\ Rev.\ D {\bf 87}, 053007 (2013);  
  R.~L.~Awasthi, M.~K.~Parida and S.~Patra,
  JHEP {\bf 1308}, 122 (2013); 
  P.~Pritimita, N.~Dash and S.~Patra,
  JHEP {\bf 1610}, 147 (2016); 
  N.~Haba, H.~Ishida and Y.~Yamaguchi,
  JHEP {\bf 1611}, 003 (2016). 


\bibitem{deGouvea:2013zba} 
  A.~de Gouvea and P.~Vogel,
  Prog.\ Part.\ Nucl.\ Phys.\  {\bf 71}, 75 (2013). 

\bibitem{Khachatryan:2015gha} 
  V.~Khachatryan {\it et al.}, 
  Phys.\ Lett.\ B {\bf 748}, 144 (2015); 
  G.~Aad {\it et al.}, 
  JHEP {\bf 1507}, 162 (2015).   

\bibitem{Dev:2013wba} 
  P.~S.~B.~Dev, A.~Pilaftsis and U.~K.~Yang,
  Phys.\ Rev.\ Lett.\  {\bf 112}, 081801 (2014); 
  D.~Alva, T.~Han and R.~Ruiz,
  JHEP {\bf 1502}, 072 (2015); 
  A.~Das and N.~Okada,
  Phys.\ Rev.\ D {\bf 93}, 033003 (2016); 
%
  C.~Degrande, O.~Mattelaer, R.~Ruiz and J.~Turner,
  Phys.\ Rev.\ D {\bf 94}, 053002 (2016). 




\bibitem{Dev:2015pga} 
  P.~S.~B.~Dev and R.~N.~Mohapatra,
  Phys.\ Rev.\ Lett.\  {\bf 115}, 181803 (2015); 
  J.~Gluza, T.~Jeli\'{n}ski and R.~Szafron,
  Phys.\ Rev.\ D {\bf 93}, 113017 (2016);
  G.~Anamiati, M.~Hirsch and E.~Nardi,
  JHEP {\bf 1610}, 010.

\bibitem{Khachatryan:2014dka} 
  V.~Khachatryan {\it et al.} [CMS Collaboration],
  Eur.\ Phys.\ J.\ C {\bf 74}, 3149 (2014). 

\bibitem{delAguila:2008cj} 
  F.~del Aguila and J.~A.~Aguilar-Saavedra,
Phys.\ Lett.\ B {\bf 672}, 158 (2009); 
  C.~Y.~Chen and P.~S.~B.~Dev,
  Phys.\ Rev.\ D {\bf 85}, 093018 (2012); 
  A.~Das and N.~Okada,
  Phys.\ Rev.\ D {\bf 88}, 113001 (2013); 
  A.~Das, P.~S.~B.~Dev and N.~Okada,
  Phys.\ Lett.\ B {\bf 735}, 364 (2014). 

\bibitem{Dev:2012zg}  
P.~S.~B.~Dev, R.~Franceschini and R.~N.~Mohapatra,
  Phys.\ Rev.\ D {\bf 86}, 093010 (2012); 
  C.~G.~Cely, A.~Ibarra, E.~Molinaro and S.~T.~Petcov,
  Phys.\ Lett.\ B {\bf 718}, 957 (2013). 

\bibitem{Blondel:2014bra} 
  A.~Blondel {\it et al.}, 
  arXiv:1411.5230 [hep-ex]; 
  E.~Izaguirre and B.~Shuve,
  Phys.\ Rev.\ D {\bf 91}, 093010 (2015); 
  C.~O.~Dib, C.~S.~Kim, K.~Wang and J.~Zhang,
  Phys.\ Rev.\ D {\bf 94}, 013005 (2016). 

\bibitem{Helo:2013esa} 
  J.~C.~Helo, M.~Hirsch and S.~Kovalenko,
  Phys.\ Rev.\ D {\bf 89}, 073005 (2014); 
  S.~Antusch, E.~Cazzato and O.~Fischer,
  JHEP {\bf 1612}, 007 (2016); 
  L.~Duarte, J.~Peressutti and O.~A.~Sampayo,
  arXiv:1610.03894.


\bibitem{Aaij:2014aba} 
  R.~Aaij {\it et al.} [LHCb Collaboration],
  Phys.\ Rev.\ Lett.\  {\bf 112}, 131802 (2014).  

\bibitem{Alekhin:2015byh} 
  S.~Alekhin {\it et al.},
  Rept.\ Prog.\ Phys.\  {\bf 79}, no. 12, 124201 (2016). 

\bibitem{Antusch:2015mia} 
  S.~Antusch and O.~Fischer,
  JHEP {\bf 1505}, 053 (2015); 
%
%
%
  S.~Banerjee {\it et al.},
  Phys.\ Rev.\ D {\bf 92}, 075002 (2015); 
  S.~Antusch, E.~Cazzato and O.~Fischer,
  arXiv:1612.02728 [hep-ph].

\bibitem{Mondal:2015zba} 
  S.~Mondal and S.~K.~Rai,
  Phys.\ Rev.\ D {\bf 93}, 011702 (2016); 
  Phys.\ Rev.\ D {\bf 94}, 033008 (2016); 
  M.~Lindner, F.~S.~Queiroz, W.~Rodejohann and C.~E.~Yaguna,
  JHEP {\bf 1606}, 140 (2016).   

\bibitem{Atre:2009rg} 
  A.~Atre, T.~Han, S.~Pascoli and B.~Zhang,
  JHEP {\bf 0905}, 030 (2009); 
  A.~de Gouvea and A.~Kobach,
  Phys.\ Rev.\ D{\bf 93}, 033005 (2016); 
%
  M.~Drewes, B.~Garbrecht, D.~Gueter and J.~Klaric,
  arXiv:1609.09069. 








\bibitem{ATLAS:2014kca} 
  G.~Aad {\it et al.}, 
  JHEP {\bf 1503}, 041 (2015); 
  CMS Collaboration, 
  CMS-PAS-HIG-14-039.

\bibitem{Dev:2016dja} 
  P.~S.~B.~Dev, R.~N.~Mohapatra and Y.~Zhang,
  JHEP {\bf 1605}, 174 (2016); 
  M.~Mitra, S.~Niyogi and M.~Spannowsky,
  arXiv:1611.09594 [hep-ph].

\bibitem{Aad:2015cxa} 
  G.~Aad {\it et al.}, 
  Phys.\ Rev.\ D {\bf 92}, 032001 (2015); 
  CMS Collaboration, 
  CMS-PAS-EXO-16-002.







\end{thebibliography}
\end{document}